\documentclass[pra,twocolumn,superscriptaddress,showpacs,preprintnumbers, 
amsmath,amssymb]{revtex4}

\usepackage{graphicx}
\def\be{\begin{equation}}
\def\ee{\end{equation}}
\def\ba{\begin{eqnarray}}
\def\ea{\end{eqnarray}}
\def\bw{\begin{widetext}}
\def\ew{\end{widetext}}

\begin{document}

\title{An efficient method for scattering problems in open billiards: Theory 
and applications.}

\author{Gursoy B. Akguc,Thomas H. Seligman}
\affiliation{Centro de Ciencias F\'isicas, Universidad Nacional
Aut\'onoma de M\'exico, A.P 48-3, 62210 Cuernavaca Mor., M\'exico}

\begin{abstract}
We present an efficient method to solve scattering problems in two-dimensional open 
billiards with two leads 
and a complicated scattering region. The basic idea is to transform the 
scattering region to a rectangle, which will lead to complicated dynamics in 
the interior, but simple boundary conditions. The method can be specialized 
to closed billiards, and it allows the treatment of interacting particles 
in the billiard. We apply this method to quantum echoes measured 
recently in a microwave cavity, and indicate, how it can be used for interacting particles.
\end{abstract}

\maketitle
\section{Introduction}
The role of two-dimensional billiards in the development
of chaos theory in general and of wave chaos in particular is
basic. The pioneering work of
Sinai \cite{sinai} 
has opened the door for a large field of research where work of Bunimovich 
and others 
\cite{bunimovich, bunimovich1} has provided us with a number of interesting shapes, for 
which chaoticity 
(at least ergodicity and the Kolmogorov property) have been proven. This 
has become 
particularly relevant as 
for soft potentials, at this point, no such example was found. There is 
even a wide-spread 
belief, that no example may exist. The field got additional impulse, from 
the fact that in scattering 
systems hyperbolicity may be easily proven
{\it e.g.} for the two and three disk problem \cite{Eckhardt}, 
and for the latter there exists 
a bounded manifold, that is chaotic. For scattering systems this property is also 
achievable for soft potentials 
\cite{jung}. The study of billiards has also proven interesting in 
mixed systems, both of the 
bound ({\it e.g.} \cite{bmushroom1, bmushroom2}) and open type \cite{mejia, 
friedrich}.

In physics, billiards have acquired increasing importance as they are
seen to represent relevant aspects of systems as small as quantum dots
\cite{qdots} and as large as planetary rings \cite{benet}. Wave 
realizations of such objects have become even more
popular, as experiments with flat microwave cavities become available from 
an ever larger number of labs 
\cite{stockmann1,stockmann2,richter1,richter2,richter3,
smilansky, sirko, anlage, Shridar,legrand}.
 Such experiments emulate single particle properties of
mesoscopic devices. Furthermore such billiards provide a very direct and 
experimentally accessible 
way to test hypothesis on classical-quantum connections,
such as the relation between the classical ensembles of random matrices 
\cite{cartan} and chaotic 
evolution, for spectra \cite{casati, bohigas, berry, leyvraz, braun}, 
wave-functions \cite{porter, gorin} 
and scattering properties \cite{seligman,vwz}. 
\par
Microwave billiards emulate the wave behavior of the corresponding 
classical billiard and thus they 
form the most direct way to test hypotheses proposed 
for classical-quantum or ray-wave connections. For closed billiards 
extremely efficient algorithms are exist
\cite{vergini, prosen}, while for scattering situations only some standard 
finite element programs are available to  
solve general problems. Recently a new technique has been proposed for 
special situations, such as 
the ripple channel \cite{Akguc1, Akguc2}. The purpose of the present paper is to
generalize this 
technique such that it 
can bee applied to typical open billiards in use as quantum dots or in 
microwave experiments. The 
basic idea is to transform the complicated border to a simple rectangle 
with openings. The price we pay is that we now have to deal with
 a much more complicated Hamiltonian instead of the simple Laplace-Beltrami 
operator. 

The advantage of this procedure thus clearly results from the simple boundary conditions.
While these also play a role for bound systems they are more pronounced in 
scattering systems. 
If we use R-matrix theory we have to implement Dirichlet and Neumann conditions in different parts 
of the surface simultaneously, and this becomes rather easy with the simple boundary conditions. 
The more complicated dynamics can be handled by in such a way, that the main numerical effort
is reduced to fast Fourier transforms, which are known to be extremely efficient.

We apply our method to obtain transmission and the power spectrum for a scattering configuration 
with mixed hilbertspace in the interior of the cavity very similar to the one 
for which scattering experiments 
have been performed successfully in Darmstadt with a superconducting microwave cavity \cite{friedrich}.
We shall refer to this configuration as Darmstadt experiment configuration (or billiard) and 
we will show
that we can reproduce the main feature, namely the scattering echo \cite{mejia}, which was 
theoretically predicted. 
Also we find, that this kind of dynamics introduces a step-structure in total transmission,   
which should be observable in microwave billiards, though total 
cross-section measurements over many channels are always
difficult.
Furthermore we show, that two-body interactions can be introduced efficiently, thus opening the 
door for a plethora of additional important applications.

We shall proceed in the next section to recall some elements of S- and 
R-matrix 
theory. Furthermore we shall connect these to concepts such as reflection 
and transmission, much 
more common in the language of mesoscopics. In the section III we attack 
the main problem, namely 
to show how we transform a complicated shape to a rectangle and what 
operator will appear in 
calculation of the the R-matrix. We then obtain a general result with a 
well defined numerical recipe, which is 
mainly based on fast Fourier transforms (FFT).
In section four we apply this technique to the Darmstadt experiment billiard used in 
\cite{friedrich} to experimentally 
confirm the existence of scattering echoes due to large integrable islands 
in the classical system. 
Note that the original theoretical prediction for a quantum system was made generically 
using a 
simple delta-kicked one-dimensional 
system, and the authors there claim, that a computation of the scattering 
billiard is not feasible to sufficiently 
short wave length with available means\cite{mejia}. 
In the following section we open the door to further applications as we discuss, how 
the method can be applied for systems with two-body interactions. 
We then give some brief conclusions.

\section{Scattering in open billiards}

\begin{figure}
\includegraphics[width=0.6\columnwidth]{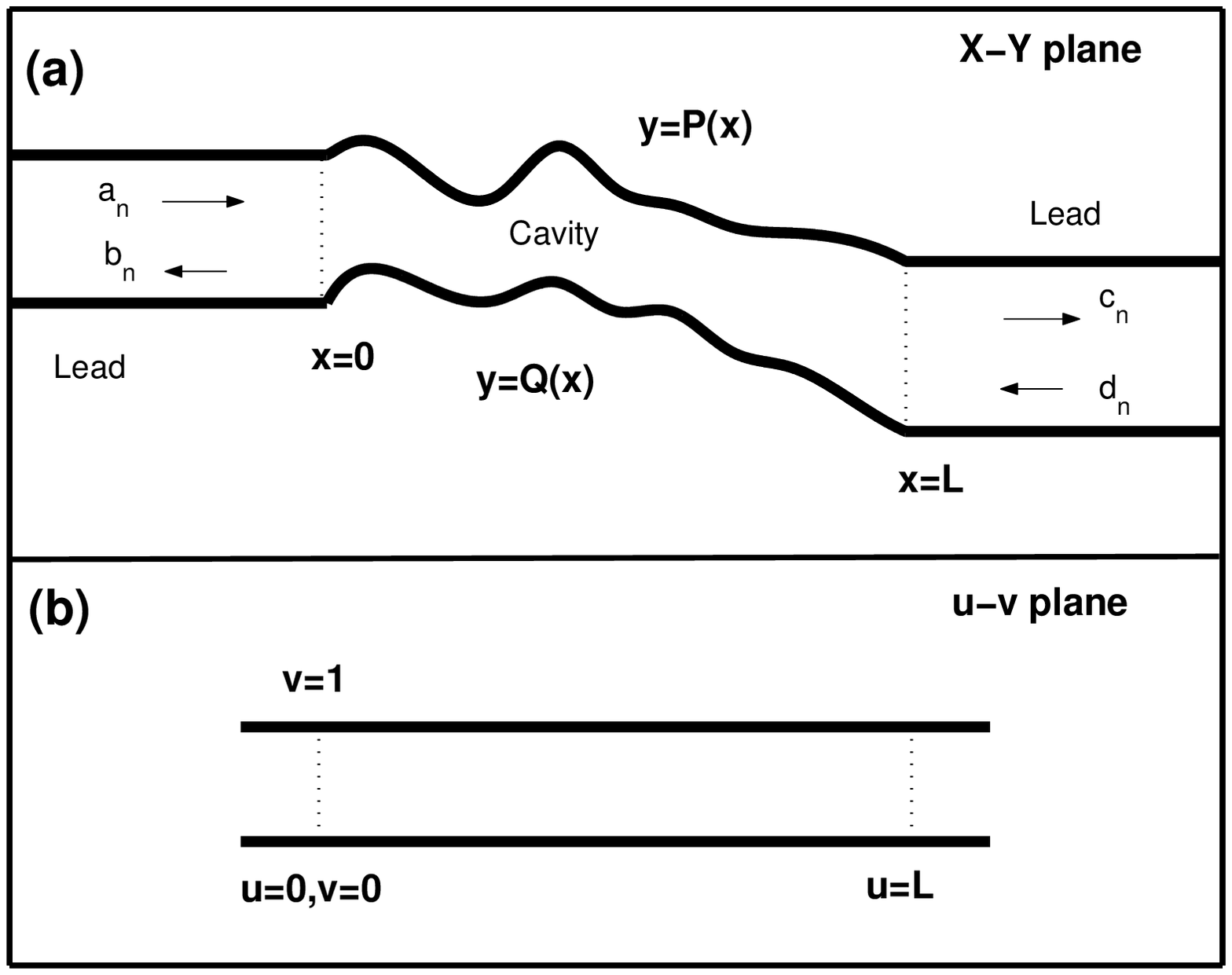} 
\caption{(a) An open billiard of length L . (b) the scattering cavity is 
transformed to a rectangular region in u-v plane}
\label{fig:guides}
\end{figure}

The system we are concerned with is sketched in Fig. \ref{fig:guides} It 
consist of a flat wave guide 
of width $w$ that supports $N$ open modes (or channels)  with a rather 
general  scattering  region  
of length $L$.  The specific method we develop, requires two leads  
(openings are in principle also possible) 
at opposing ends of a scattering region. We shall arbitrarily fix them 
to be on the right and on the left, as in the figure. The scattering 
region between n the leads or openings shall be described by two single-valued 
functions of a coordinate say $x$  defined between the two leads.
No additional scatterers in the so defined cavity are allowed.
We will assume, that the leads have equal width. The latter is not 
essential for the method but simplifies notation.
The scattering problem will be defined in terms of 
 a $2N\times 
2N$ scattering matrix $S$. Having nano structures in mind we can rewrite the 
S-matrix as 

\begin{equation}
S = \left( 
\begin{array}{cc}
r & t' \\ t & r' 
\end{array} 
\right), 
\end{equation}
where $r$ ($r'$) and $t$ ($t'$) are the reflection and transmission matrices 
for incidence on the left (right) 
of the scattering region. The dimensionless conductance is obtained from 
$S$ as 
\begin{equation}
T = \rm{tr}(tt^{\dagger}).
\end{equation}

Our calculations will be based on the traditional R-matrix approach 
\cite{Wigner, Lane}. 
This formalism  has been adapted to obtain the S matrix for a wave guide in 
previous works \cite{Akguc1,Akguc2,Reichl}. Until now only a restricted choices of  
scattering regions has been investigated. 
For example only one of the wall was taken of sinusoidal shape while the other one  
was kept flat (ripple 
billiard). We want to generalize this approach to a generic open two-lead
cavity without obstacles, where one can choose any unique
differentiable function of $P(x)$ and $Q(x)$ in Fig.~1 to represent the 
upper and lower walls of the scattering cavity. 

We wish to remind the reader, that  the $R$-matrix approach 
relates the S-matrix to a matrix $R $ determined by  boundary conditions relating internal 
and external functions at the edges, where the leads are connected to the 
scattering region, such that
\be
\psi |_{\Gamma}=R \nabla \psi |_{\Gamma}
\ee
at $x=\Gamma$ as in of the boundaries represented as dotted lines in Fig.~1. $R$-matrix can be generalized for number of lead connection and number of modes in the leads. $R$-matrix can be related to $S$-matrix,
\be
S= \frac{1-ikR}{1+ikR} 
\ee

Rather than recalling details of $R$-matrix theory we 
will evoke the spirit of the method by presenting a trivial one dimensional 
example:

In Fig.~2 we show the 1D example we want to solve using $R$-matrix method.
In this example we have 2 asymptotes as in the general case we discuss but the
$r$ and $t$ are functions instead of a matrices as in the 1 mode case for the real
cavity problems. 
 The exact $S$-matrix can be found in this case and the transmission probability is given by
\be
T_{exact}(E)=\frac{1}{(1+(\frac{V_0^2}{(E(E-V_0))(2 \sin(k))^2})}
\ee
for a constant potential step of height $V_0$ and length 1.
 
Numerically we use a basis for the Reaction region,(R in Fig.~2), 
which is given by $cos(m\pi x), m=0,1,\dots \infty$. $x_l=0$ and $x_r=1$ 
have been chosen. 
Using this basis, the R-matrix elements are
given by,
\ba
R_{rr}=\frac{1}{E-V_0}+\sum_{m=1}^{\infty} \frac{2}{E-m^2\pi^2-V_0}=R_{ll} 
\nonumber \\
R_{rl}=\frac{1}{E-V_0}+\sum_{m=1}^{\infty} \frac{2 \cos(m 
\pi)}{E-m^2\pi^2-V_0}=R_{lr}. \nonumber \\
\ea
This series should be truncated at some finite value for a numerical calculation; 
$m=1000$ is used in 
Fig.~2. The S matrix is obtained from R matrix as
\be
S = \left( 
\begin{array}{cc}
1 & 0 \\ 0 & e^{-ik} 
\end{array} 
\right)
\frac{ \left( 
\begin{array}{cc}
1 & 0 \\ 0 & 1 
\end{array} 
\right)-ik \left( 
\begin{array}{cc}
R_{rr} & R_{rl} \\ R_{lr} & R_{ll} 
\end{array} 
\right)}{\left( 
\begin{array}{cc}
1 & 0 \\ 0 & 1 
\end{array} 
\right)+ik \left( 
\begin{array}{cc}
R_{rr} & R_{rl} \\ R_{lr} & R_{ll} 
\end{array} 
\right)}
 \left( 
\begin{array}{cc}
1 & 0 \\ 0 & e^{-ik} 
\end{array} 
\right)
\nonumber
\ee
Using this S matrix we plot the transmission probability ,$|S_{1,2}|^2$ in 
Fig.~1. We note that generalizing the potential in this example to any
shape can be done by using a finite difference method.\cite{Akguc3} 

\begin{figure}
\includegraphics[width=0.6\columnwidth]{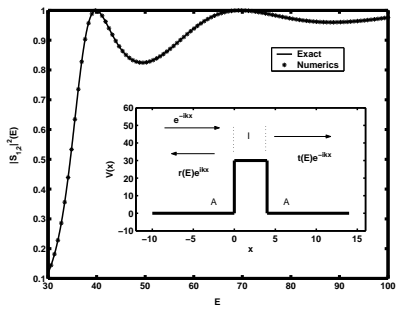} 
\caption{Scattering from a 1D barrier. The solid line shows the transmission 
probability as a function of energy and stars 
are the numerical calculations. In the inset the geometry of the scattering system is
shown. $I$ indicates the interaction region, $A$ the asymptotic regions.}
\label{fig:simple} 
\end{figure}

The method to obtain the S-Matrix thus proceeds in two steps: First obtain 
the solution of the boundary value problem inside 
the scattering region with Neumann and Dirichlet boundary conditions at 
lead connections and cavity walls respectively. Second 
connect the known solution in the 
asymptotic region(leads) to the one obtained for the scattering region by 
imposing continuity of the solution.  In the method we present 
surprisingly  the  
most time consuming step in numerics will be to obtain a complete set of function to expand 
the scattering 
wave function inside the cavity region, which is
a prerequisite to the first step. 

\section{The Method}

We shall thus start by presenting a very efficient method to deal with the 
problem inside the scattering region. This method can also be applied to 
closed cavities, as we will point out further down. As we will use 
transformed coordinates the second step will not be entirely trivial and 
will be presented once we have the internal solutions.

\subsection{Cavity Region} 
We want to solve the Schr\"odinger equation in the cavity with specific 
boundary 
conditions, namely Neumann conditions at the open boundaries, and 
Dirichlet conditions at the walls of the billiard. With other words, 
the derivative of the wave function 
is set to be zero at the dotted line in Fig.~1 and the wave function itself is 
set to zero at surface of the billiard. Thus we have the equation
\be
H|\Psi(x,y)>=-\frac{\hbar^2}{2m}(\nabla_x^2 
+\nabla_y^2)\Psi(x,y)=E_n\Psi(x,y),
\ee
with the boundary conditions 
$\nabla_x\Psi(x,y)|_{x=0}=0$, $\nabla_x\Psi(x,y)|_{x=L}=0$, 
$\Psi(x,y)|_{y=P(x)}=0$ and  $\Psi(x,y)|_{y=Q(x)}=0$. 
The eigenvalues and eigenvectors of this equation are used to represent 
the 
scattering dynamics inside the cavity.

The solution of the Schr\"odinger equation in the coordinates $x,y$ is 
complicated due to the arbitrary shape of the boundaries so we 
use following transformation to new coordinates $u,v$. 
\ba
u&=&x \nonumber \\
v&=&\frac{y-Q(x)}{P(x)-Q(x)}
\ea 
This change of variable transforms the complicated boundary to a simple 
one,{\it i.e.} a rectangular region. The price we pay, is that the form 
of 
the Schroedinger equation becomes more
complicated. The Hamiltonian operator in the $u-v$ plane can be obtained by 
using 
the above transformation and 
complete derivatives. The final result is 
\ba
H = \nabla_u^2+\left( \frac{1+(Q_u+vJ_u)^2}{J^2} \right)\nabla_v^2-2\left( 
\frac{Q_u+vJ_u}{J}\right) \nabla_{u,v} \nonumber \\
-\left( \frac{Q_{u,u}+vJ_{u,u}}{J}\right)\nabla_v+2\left( 
\frac{(Q_u+vJ_u)J_u}{J^2}\right)\nabla_v
\ea
where $J$ is defined as $J=P-Q$ and subscripts indicate partial 
derivatives. In this form the symmetric 
nature of Hamiltonian is not obvious. Therefore we will
represent the same equation in the following matrix form,\cite{dewit}
\be
H = \frac{1}{J}\left(\nabla_{\alpha} J 
g^{\alpha,\beta}\nabla_{\beta}\right) \hspace{10pt} \alpha=u,v \hspace{5pt} 
\beta=u,v
\ee
where $g$ is a metric and equivalent to,
\be
g\equiv \left(
\begin{array}{cc}
1 &\frac{ -(Q_u+vJ_u)}{J}\\
\frac{-(Q_u+vJ_u)}{J} & \frac{1+(Q_u+vJ_u)^2}{J^2}
\end{array}
\right)
\ee
The meaning of $J$ becomes clear when we notice it is related to the 
determinant of the metric. It plays the role of Jacobian of the transformation 
from the x-y to the u-v plane. The wave function in the u-v plane can be expanded 
in terms of a basis 
\ba
\Psi_k(u,v)&=&\sum_{n,m} B_{n,m}^k \psi_{n,m} \nonumber \\
&=&\sum_{n,m} B_{n,m}^k\frac{1}{\sqrt{J}}\frac{2}{\sqrt{L}}\cos(\frac{m\pi 
u}{L})\sin(n\pi v)
\ea
which satisfies the boundary conditions at $u=0,u=L$ and $v=0,v=1$ 
automatically.
To determine the unknown coefficients $B_{n,m}^k$ we will use the 
orthanormality 
condition. In the original coordinates plane 
waves are orthonormal; in terms of $u, v$ they must be orthonormal with a 
weight function given by the metric
\be
<\Psi |\Psi>\equiv \int \int dudv J  
\Psi_{k}(u,v)\Psi_{k'}(u,v)=\delta_{k,k'}
\ee
After multiplying the eigenvalue equation with another eigenfunction and 
integrating
we obtain the matrix equation, 
\be
\sum_{l}\sum_{l'} H_{l l'}B_{l'}^k=B_{l}^k
\ee
where we use new indices such that $(n,m) \rightarrow l$ and $(n',m') 
\rightarrow 
l'$. Eigenvalues $E_k$ correspond to energy and eigenvectors contain 
the unknown coefficients $B_{l}^k$ of the
wavefunction. Using the matrix form of $H$ we have,
\be
\int \int dudv J \psi_{l} \frac{1}{J} \nabla_{\alpha} J g^{\alpha 
\beta}\nabla_{\beta} \psi_{l'} =E_k\delta_{l,l'}.
\ee
We can integrate by parts. The surface terms are zero due to the boundary 
condition we have chosen.  
We thus find
\be
-\int \nabla_{\alpha}\psi_l J g^{\alpha \beta} \nabla_{\beta} 
\psi_{l'}=E_k \delta_{l,l'}.
\label{eq:mateq}
\ee
Since the metric is symmetric, the matrix obtained from this equation is real 
symmetric 
and positive definite which guarantees positive energy eigenvalues. 

Once the eigenvalue equation is solved one can obtain scattering wave 
function 
for any energy within the range of wave length we allow. This 
makes the method efficient compared to direct integration of the 
differential equation by some mesh based method like finite element 
\cite{Akguc1}. 
The cut off on the number
of basis states determines the accuracy and range of validity of the 
method. 
In practice it is necessary to decide how many modes 
to choose in $u$ and $v$ direction. A corresponding integrable geometry 
gives an idea about 
these numbers. Also one needs to choose only the lower part of the 
eigenvalue 
spectrum in the bound region to ensure that
truncation wont bring errors due to poor solutions of the bounded 
problem. 
For example $m=90$ , $n=50$ and using the first 3000 eigenvalues 
of the 4500X4500 matrix is suitable for a rectangular region with length 
twice 
its width. The limit of the highest wavelength that can be attained in leads is
connected to the lead size to the scattering size. If the size of scattering region is much bigger than the size of the leads than the method is limited to only couple of modes in leads due to the cut off on the number of eigenvalues that has to be done in cavity region. In the application we discuss in this paper 
this is not the case since size of cavity is smaller than leads.

  We need an efficient method to calculate the matrix elements. 
Each matrix elements is a double integral over u and v. Due to the form of 
the metric and sinusoidal nature 
of the basis states we can achieve this writing the terms of Eq.~\ref{eq:mateq} 
as follows:
\ba
&<&\Psi|H|\Psi>=\int du \nabla_u (\frac{C_m}{\sqrt{J}}) J \nabla_u 
(\frac{C_{m'}}{\sqrt{J}}) \cdot \int dv S_n S_{n'} \nonumber \\
&+& n'\pi \int du  \nabla_u (\frac{C_m}{\sqrt{J}}) (-Q_u) 
\frac{C_{m'}}{\sqrt{J}} \cdot \int dv S_n C_{n'} \nonumber \\
&+& n'\pi \int du  \nabla_u (\frac{C_m}{\sqrt{J}}) (-J_u) 
\frac{C_{m'}}{\sqrt{J}} \cdot \int dv S_n v C_{n'} \nonumber \\
&+& n\pi \int du \frac{C_m}{\sqrt{J}} (-Q_u) \frac{C_{m'}}{\sqrt{J}} \cdot 
\int dv C_n S_{n'} \nonumber \\
&+& n\pi \int du \frac{C_m}{\sqrt{J}} (-J_u) \frac{C_{m'}}{\sqrt{J}} \cdot 
\int dv C_n v S_{n'} \nonumber \\
&+& n n'\pi^2 \int du C_m C_{m'} \frac{(1+Q_u^2)}{J^2} \cdot \int dv C_n 
C_{n'} \nonumber \\
&+& n n'\pi^2 \int du C_m C_{m'} \frac{(2Q_uJ_u)}{J^2} \cdot \int dv C_n v 
C_{n'} \nonumber \\
&+&n n' \pi^2 \int du C_m C_{m'} \frac{(J_u^2)}{J^2} \cdot \int dv C_n v^2 
C_{n'}. \nonumber 
\ea
Here we use the abbreviations $C_m \rightarrow \cos(m \pi u/L)$, $S_n 
\rightarrow \sin(n \pi v)$ and 
$C_n \rightarrow \cos(n \pi v)$. The integral over $v$ can be performed 
analytically. 
All the terms except the first one
are of the form of a function multiplying a $C_m$ term. These integrals can 
be 
done using a fast Fourier 
transform. The first term can be put in the same form, after some algebra:
\ba
K_1=(\frac{m^2 \pi^2 }{L^2}\delta_{m,m'} -\int du C_m \nabla_u(C_m') 
\nabla_u(\log(J)/2) \nonumber \\
-\int du C_{m'} \nabla_u (C_m) \nabla_u(\log(J)/2) \nonumber \\
+\int du C_m C_{m'} (\nabla_u(\log(J)/2))^2) \nonumber
\ea

Therefore, when the functions $P$, $Q$ and $J$ are given as an array of 
dimension 
$M$, the integral is equal  to
\be
\int du C_m F(u)=\frac{L}{M} Re(FFT(F)_{m-1})
\ee
{\it i.e.} the $m-1$th element of the real part of the Fourier transform 
of 
the function is equal to the integral of the same 
function multiplied by $C_m$ . Note that one needs to incorporate end 
point correction \cite{Numrec} 
to be more accurate; one way is to extend integration function 
symmetrically at the end point and halve the result. 
This is very efficient since each array of integrals is 
calculated by a single Fourier transform. A total of seven Fourier transform 
is enough to calculate 
all integrals for this problem. The overall matrix elements can be written 
using the result of 
exact integration in v direction as
\ba
<\Psi|H|\Psi>=\delta_{n,n'}(\frac{(m-1)^2 \pi^2}{L^2}- F_1(m,m')\nonumber 
\\
-F_1(m',m)+F_2(m,m')) \nonumber \\
+D_1(n',n) F_3(m',m)+D_2(n,n') F_4(m',m) \nonumber \\
+D_1(n,n') F_3(m,m')+D_2(n,n') F_4(m,m')\nonumber \\
+D_3(n,n') F_5(m,m')+D_4(n,n') F_6(m,m') \nonumber \\
+D_5(n,n') F_7(m,m'),
\ea
where $v$ integrals are obtained analytically, 
\ba
D_1(n,n')=(1-\delta_{n,n'}) n' n \left( 
\frac{(-1+(-1)^{(n+n')}}{(n'^2-n^2)}\right) \nonumber \\
D_2(n,n')=\frac{-\delta_{n,n'}}{4}+ (1-\delta_{n,n'}) n n' \left( 
\frac{-1^{(n+n')}}{(n'^2-n^2)}\right) \nonumber \\
D_3(n,n')=\delta_{n,n'} (n^2 \pi^2/2) \nonumber \\
D_4(n,n')=\delta_{n,n'} (n^2 \pi^2/4)+ \nonumber \\
(1-\delta_{n,n'})n n' 
\left(\frac{(-1+(-1)^{(nn-n)})(n^2+n'^2)}{(n'^2-n^2))^2}\right) \nonumber 
\\ 
D_5(n,n')=\delta_{n,n'} (n^2 \pi^2/2)(1/6+1/(4 n^2 \pi^2))+ \nonumber \\
(1-\delta_{n,n'})n n' \left(\frac{2 
(-1)^{(nn-n)}(n^2+n'^2)}{(n'^2-n^2))^2}\right) \nonumber \\
\ea 
and $u$ integrals 
\ba
F_1(m,m')=\frac{L}{M} Re(FFT_m(\nabla_u(C_{m'}) \nabla_u(log(J/2)))) 
\nonumber \\
F_2(m,m')=\frac{L}{M} Re(FFT_m(C_{m'} \nabla_u(log(J/2)))) \nonumber \\ 
F_3(m,m')=\frac{L}{M} Re(FFT_m(-Q_u \nabla_u(\frac{C_{m'}}{\sqrt{J}}))) 
\nonumber \\ 
F_4(m,m')=\frac{L}{M} Re(FFT_m(-J_u \nabla_u(\frac{C_{m'}}{\sqrt{J}}))) 
\nonumber \\  
F_5(m,m')=\frac{L}{M} Re(FFT_m(\frac{( C_{m'}+C_{m'}Q_u^2)}{J^2})) 
\nonumber \\ 
F_6(m,m')=\frac{L}{M} Re(FFT_m(2 C_{m'} \frac{Q_u J_u}{J^2} )) \nonumber 
\\ 
F_7(m,m')=\frac{L}{M} Re(FFT_m(C_{m'}\frac{J_u^2}{J^2} )) \nonumber \\ 
\ea

The final step is to find eigenvalues and eigenvectors of this matrix 
which in turn gives yield the unknown expansion 
coefficients of the wave function in the u-v plane and obviously the energy eigenvalues.
 It is possible to obtain 
the wave function as functions of $x,y$ applying the inverse transform to 
the wave function in in terms of $u,v$. Since the expansion  
coefficients are already determined this transformation is straight forward. 
%%%%%%%%%%%%%%%%%%%
\subsection{Leads and Scattering Matrix}
With the knowledge of states in the cavity region it is possible to 
couple 
known asymptotic, such as leads to the cavity. The corresponding solution 
on the left and on the right leads to
\ba
\Psi_{L}^n=\left( \frac{a_n}{\sqrt{k_n}}e^{ik_nx} 
-\frac{b_n}{\sqrt{k_n}}e^{-ik_nx}\right)\sin(\frac{n\pi y}{w}) \nonumber \\
\Psi_{R}^n=\left( 
\frac{c_n}{\sqrt{k_n}}e^{-ik_nx}-\frac{d_n}{\sqrt{k_n}}e^{ik_nx}\right)\sin(\frac{n\pi 
y}{w}) \nonumber \\
\ea
where $w$ is the lead width, $a_n$, $b_n$,$c_n$ and, $d_n$ are scattering 
amplitudes and the wave vector is given by 
\be 
k_n=\sqrt{\left( \frac{n\pi}{w} \right)^2-\frac{2mE}{\hbar^2}}.
\ee
We here assume propagating modes but it is possible to add evanescent 
modes 
with complex k vectors to the calculation. 
Their effect is discussed in ref.~\cite{Akguc2}
 
As shown in references \cite{Akguc1}\cite{Akguc2}\cite{Reichl} the 
energy eigenfuntions $|E>$ of the total scattering system have contributions 
from both cavity and leads.
\be
<x,y|E>=\sum_{j=1}^{\infty} \gamma_j\phi_j(x,y) + 
\sum_{n=1}^{\infty}(\Gamma_n^L \Psi_L^n+ \Gamma_n^R \Psi_R^n)
\ee
where $\phi_j$ are basis states calculated for the cavity alone. 

The continuity of scattering wave functions at the lead boundary gives the
connection between the leads and the cavity region:
\ba
\Psi_n^{\alpha}=\sum_{n'=1}^{\infty}R_{\alpha 
L}(n,n')\nabla_x\Psi_n^{\alpha}|_{x_L} \nonumber \\
-\sum_{n'=1}^{\infty}R_{\alpha R}(n,n')\nabla_x\Psi_n^{\alpha}|_{x_R} 
\nonumber \\  
\ea

where

\be
R_{\alpha 
\beta}(n,n')=\frac{\hbar^2}{2m}\sum_{j=1}^{\infty}\frac{\phi_{j,n}(x_{\alpha})\phi_{j,n'}(x_{\beta})}{E-\gamma_j}
\ee

is the (n,n')th  matrix element of the R matrix. 
Here $\phi_{j,n}$ is the overlap between cavity states and lead 
functions and given by
\be
\phi_{j,n}(x_{\alpha})=\sqrt{\frac{2}{L}}\int_0^{\infty}dy 
\phi_j(x_{\alpha})\sin(\frac{n \pi y}{w})
\ee
where $ \alpha \equiv L,R$ and $x_L=0$, $X_R=L$ in Fig.~1.

\section{Application: Quantum echoes from cavities with a large integrable 
island}

As an application of our method we solve the scattering problem in the 
Darmstadt-experiment geometry shown in Fig.~3. 
A similar geometry to Fig.~3(a) has been used in recent experiments 
\cite{friedrich}. The analytical curves of upper 
and lower wall in Fig.~3a(a) are given as follows, 
\be
P(x)=\lambda \exp (-\frac{\alpha x^2}{\lambda^2}) \hspace{.5 cm} and 
\hspace{.5 cm} Q(x)=\lambda \left( \beta-\frac{\gamma x^2}{\lambda^2} 
\right).
\ee
We use the same parameters as in experiment such as $\alpha=0.161$, $\beta 
=0.2$ and $\gamma=0.1$. 
While in experiment a scaling factor of $\lambda=5cm$ is used we use 
$\lambda=0.432$ in arbitrary units and we shift 
the origin of the coordinates to the beginning of the left lead for 
convenience.Note though, that the original experiment does not connect to leads!
Rather the opening is arbitrary, though in fact naturally finite. We will return 
to these differences later.

Apart from the original geometry we also consider a geometry with some small 
irregularity in the center of the cavity. Such an irregularity should 
destroy the stability of the central periodic orbit in the classical system
and thus at least damage or deform the large regular island significantly.
This corresponds to the mentions introduction of a perturbing scatterer 
in the experiment, which destroyed the echo \cite{friedrich}. Furthermore we 
consider that the entire parabolic wall may perturbed in a disordered way.
\begin{figure}
\includegraphics[width=0.6\columnwidth]{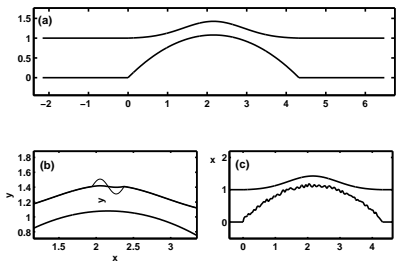} 
\caption{(a) The Darmstadt-experiment geometry of length $L=4.32$ with leads 
$W=1$. (b)same as in (a) but with
some small disorder in the middle of the billiard. (c) The lower wall in (a) 
has been  replaced by a disordered wall.}
\label{fig:eachog}
\end{figure}

After calculating the  S matrices for these systems we found the total conductance 
as described in section 1. 
The results for the 3 different geometries can be seen in Fig.~4. We see 
that for the geometry used 
in experiment there are steps in the conductance. These steps do not 
correspond to the opening of 
channels in the leads. This is because of the smaller width of the 
scattering region. The minimum 
width in channel is $w_{min}=0.3097$ with corresponding $k_{eff}=3.23$ 
 matches 
where the first step seems to start in Fig.~4. With other words, this 
is the point where we pass from evanescent 
or tunneling transmission to open channels in the necks presented by the 
Darmstadt-experiment configuration.
\begin{figure}
\includegraphics[width=0.6\columnwidth]{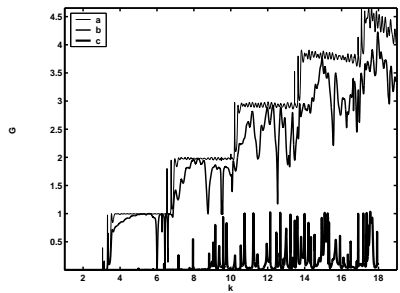} 
\caption{ Conductance versus wave vector for 3 different geometries 
corresponding to (a), (b) and (c) in Fig.~3}
\label{fig:echo1}
\end{figure}
While the initial rise thus is rather trivial, the step structure is not,
as is sensitive to even a small disturbance. The solid 
line in Fig~4 shows the total 
conductance through the same geometry but with a small wiggle at the center of the cavity. 
The exact shape of the wiggle is not very relevant. We choose 
a sinusoidal perturbation of the upper wall in the form $0.01 \sin(10 \pi 
x_s/L)$ where $x_s$ is the x 
coordinate in the interval, $9L/20<x_s<11L/20$. (In Fig.~4 a 10 times 
bigger scaled version of 
perturbation is also shown to give an idea of the shape of perturbation.) 
The effect of perturbation 
is huge on conductance data. The step structure tends to be lost for 
higher energies.

We also show the result of surface disorder on conductance by the dot-dashed curve 
in Fig.4. 
We implemented surface disorder by first dividing the boundary curve 
into $n$ pieces and 
move these pieces by a random amount, $\eta$ in y direction. We apply a 
spline interpolation to 
connect the pieces to form a smooth boundary. We see that for $\eta=0.2$ 
and $n=100$ 
the step structure is lost and only resonance transmission can be seen. We 
note that a partial 
step can be observed for small disorder parameters.   
  
We calculate the length spectrum by taking a Fourier transform of the 
transmission amplitude 
\be
t_{nm}(L)=\int_{k_{min}}^{k_{max}} dk_F t_{nm} (k_F) \exp(-i k_F L).
\label{eq:spec}
\ee
This identification can be done based on a semi-classical description where 
one can sum over different 
periodic orbits to get the transmission amplitudes \cite{gutz}. In Fig.~5 we show 
this quantity for $t_{11}(L)$ over an interval $1<k_F 
<19 $  with $k$ in units of $\pi/w$.  Note that in this interval we have a 
an increasing transmission matrix size each time a channel opens. we added 
the first element of each transmission matrix to form $t_{11}$ over the 
given k interval.
\begin{figure}
\includegraphics[width=0.6\columnwidth]{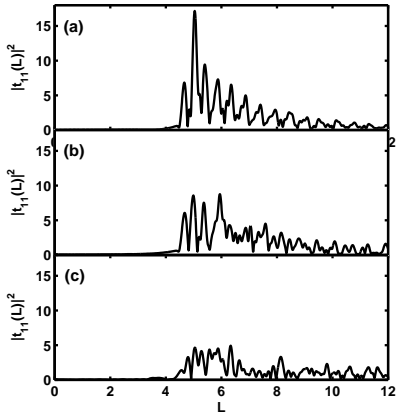} 
\caption{(a) The length spectrum for the geometry in Fig.~4(a),(b) same for the 
geometry in Fig.~4(b) (c) same for the geometry in Fig.~4(c).}
\label{fig:echo2}
\end{figure}
In Fig.~5a we see the periodic oscillations starting at $L=4.32$. Since 
length can be associated to 
time for a constant speed, such as expected in an electromagnetic cavity,
this is similar to the echos observed in experiment. 
The period seems to be constant in the data with a second harmonic 
entering at higher lengths suggesting a shortening in the period. 
In the experiment a slight shortening of the period for short times
appears.This we do not see. 
On the other hand the experiment does not show the drastic shortening 
with the higher harmonics. We mus keep in mind, that the problems are {\it not}
identical, because of the finite diameter leads we attached here. Also we
measure transmission to the entire 
channel, rather than to an antenna placed near a semi-cclassically significant manifold
as the latter is not meaningful in a billiard with leads. 
On the other hand 
effects of truncation must be studied carefully in each case. At this point we 
wish to stress, that the semi-classical structure associated with the scattering echoes again
is easily seen, and -- as in the experiment -- for wave length far from semi-classics.

To get a better impression we look at the power spectrum for the geometry in Fig.~3(a).
In Fig.~6(a) we show this spectrum calculated from $k(\pi/w)=6$ to $k(\pi/w)=9$. 
Since the minimum transmission matrix is 6X6 we sum over 6 modes. This region 
roughly corresponds to the third plateau in Fig.~4. 

\be
P=\sum_{n=1}^{n=6} \sum_{m=1}^{m=6} |t_{nm}(L)|^2,
\ee
where $t_{nm}$ is calculated using equation \ref{eq:spec} but with appropriate 
limits for k integration. In this figure we obtain peak values at $L_1=4.78$,
$L_2=6.65$, $L_3=8.05$ ,$L_4=9.1$,$L_5=10.2$. 
The peak value corresponds to the simplest direct transmission orbit starting 
at the middle of the channel opening and hitting the center of Gaussian shaped 
wall before exit. As it can be seen here period decreases
 with increasing length. We observe oscillations up to high lengths as 
can be seen in inset. 

We also show  similar calculations in the interval 
$k(\pi/w)=10 <k(\pi/w)<k(\pi/w)=13$ including 10 modes in Fig.6(b). 
Although we have different limits to the integration and different size 
of the matrix. We get results very similar to those the Fig.~6(a).

\begin{figure}
\includegraphics[width=0.6\columnwidth]{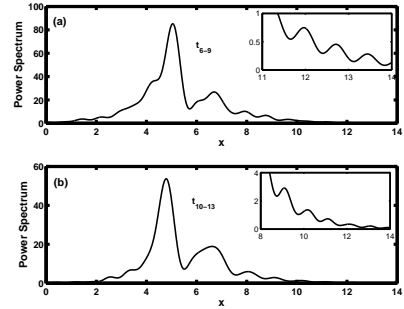} 
\caption{(a) Power spectrum from $k(\pi/d)=6$ to $k(\pi/d)=9$ including 6 modes. 
(b)Power spectrum from $k(\pi/d)=10$ to $k(\pi/d)=13$ including 10 modes. }
\label{fig:echo3}
\end{figure}

\section{Toward many-body problems}

Real naosystems have interacting particles, and treatment of such systems is incipient. 
We shall show, that the methods presented here is very promising for such situations, 
at least as long as the particles enter or leave the quantum dot only one at a time.
The basic problem he consists of treating two-body interactions, as we may typically 
encounter in a few-body system.
The two particle inside a wave guide will provide the main difficulty.
It is also an important system on its own right (see the review in \cite {Leon}).

Lets consider two particles inside a cavity region as shown in Fig.~1.
We will omit the problems associated with anti-symmetrization or symmetrization for 
Fermions or Bosons
respectively, by assuming distinguishable particles, but standard methods can be used 
to take care of the corresponding symmetry adaption as necessary.
The total Hamiltonian is,
\be
H=\nabla_{x_1}^2 +\nabla_{y_1}^2 + \nabla_{x_2}^2 +\nabla_{y_2}^2+V(|x_1-x_2|,|y_1-y_2|)
\ee
with a potential $V$ which depends on the distance of the two particles.
The interaction term can be written in terms of single particle wave functions
as follows, 
\ba
H_{ijkl}&=&\int \int dx_1 dx_2 dy_1 dy_2 \phi_i(x_1,y_1) \phi_j(x_2,y_2) \nonumber\\
 V&(|&x_1-x_2|,|y_1-y_2|)\phi_k(x_1,y_1) \phi_l(x_2,y_2)
\ea
where $\phi_i(x,y)$ is the single particle solution in the cavity. 
Energies of the interacting particle can be found after diagonalizing this matrix. 
As the form of the single particle
 functions is not simple this is not an easy calculation.
But as we discussed in this paper, we can transform this integration to
u-v plane where we know the expansion of the single particle eigenfunctions in 
terms of 
simple harmonics. In u-v coordinates the distance between particles wont be a line 
but a curve given by the metric. The matrix element in u-v coordinates becomes, 
\ba
H_{ijkl}&=&\int \int du1 du2 dv1 dv2 \phi^i(u1,v1) \phi^j(u2,v2) \nonumber \\
&V&(u1,u2,v1,v2) \phi^k(u1,v1) \phi^l(u2,v2)
\ea
where $\phi^i(u,v)=\sum_{mn}B_{m,n}^i\cos(m \pi u/L) \sin(n \pi v)$ with coefficients 
$B_{m,n}^i$  calculated  from a single particle equation.
The harmonic expansion makes the integration simple.

\section{conclusion}

We present a new method to treat open billiards, that has significant 
advantages over older techniques. It is designed for billiards with two 
leads, though generalizations are possible. The basic limitation is that 
both irregular boundaries that connect the leads be single valued 
functions of some space coordinate, that connects the center of the two 
leads. We then transformed the scattering region into a rectangle and 
calculate the quantum dynamics which become involved. The boundary 
conditions now result trivial, and this allows to calculate the R-matrix 
easily and thus solve the scattering problem. It is possible to write 
the dynamics of two or more interacting particles in this form, as 
long as we do not attempts to transform to relative and center of mass 
coordinates. This would destroy0y the simple boundary conditions. 
To show the 
power of the method we calculated the quantum scattering echoes for such a two lead 
structure very near to one actually used in a microwave experiment. The results 
of this experiment were not accessible to a numerical 
calculation with standard microwave programs, but we obtained results 
essentially consistent with experiment, though we see a secondary effects not apparent 
in the experiment. Future research will have to show whether this effect is 
suppressed in the experiment, whether it is due to the use of leads or
whether it is a consequence of the cutoff we use.
 This should not distract from the fact, 
that the main effect can be seen clearly.
Concerning the example we also note that the calculation reveals a step structure 
in the transmission, which is directly related to the scattering echoes 
and was not previously noted.

Acknowledgments

The authors are grateful for useful discussions to C. Jung. Financial support under 
DGAPA project IN191603 and CONACyT project 43375  is acknowledged.

\end{document}